\documentclass[%
 aip,
 amsmath,amssymb,
reprint,%
twocolumn
]{revtex4-1}
\usepackage {CJK}
\usepackage{graphicx}
\usepackage{dcolumn}
\usepackage{bm}
\usepackage{color}
\usepackage{hyperref}
\newcommand\We{\mbox{\textit{We}}}  
\newcommand\Oh{\mbox{\textit{Oh}}}  
\begin{document}
\begin{CJK*}{GB}{gbsn}

\title{Droplet deformation and breakup in shear flow of air}
\author{Zhikun Xu (徐志坤)}
\affiliation{State Key Laboratory of Engines, Tianjin University, Tianjin, 300072, China.}%
\author{Tianyou Wang (王天友)}
\affiliation{State Key Laboratory of Engines, Tianjin University, Tianjin, 300072, China.}%
\author{Zhizhao Che (车志钊)}
\email{chezhizhao@tju.edu.cn}
 \affiliation{State Key Laboratory of Engines, Tianjin University, Tianjin, 300072, China.}

\date{\today}

\begin{abstract}
The deformation and breakup of droplets in air flows is important in many applications of spray and atomization processes. However, the shear effect of airflow has never been reported. In this study, the deformation and breakup of droplets in the shear flow of air is investigated experimentally using high-speed imaging, digital image processing, and particle image velocimetry. We identify a new breakup mode of droplets, i.e., the butterfly breakup, in which the strong aerodynamic pressure on the lower part of the droplet leads to the deflection of the droplet and then the formation of a butterfly-shaped bag. A regime map of the droplet breakup is produced, and the transitions between different modes are obtained based on scaling analysis. The elongation and the fragmentation of the droplet rim are analyzed, and the results show that they are significantly affected by the shear via the formation and the growth of nodes on the rim.\end{abstract}


\maketitle
\end{CJK*}

\section{Introduction}\label{sec:sec01}
Liquid atomization is a common phenomenon in various fields of scientific research and engineering applications, such as combustion engines \citep{Lefebvre2017Atomization, Omidvar2019Spray, Broumand2016Liquidjet}, chemical industry \citep{Ochowiak2019SprayTower, Xue2019Gasifier}, and material preparations \citep{Fan2016ThermalBarrierCoatings}. In these atomization processes, secondary breakup, i.e., droplet deformation and breakup into fragments due to aerodynamic forces in gas streams, plays a fundamental role \citep{Guildenbecher2009SecondaryAtomization}. The secondary breakup produces numerous small droplets with a large surface-area-to-volume ratio, and has an essential influence on enhancing heat/mass transfer and increasing evaporation/reaction rate in the corresponding applications. For the breakup of individual droplets in airflows \citep{Guildenbecher2009SecondaryAtomization, Jain2015SecondaryBreakup, Theofanous2011DropBreakup, Suryaprakash2019SecondaryBreakup}, the Weber number ($\We_g$) and the Ohnesorge number ($\Oh$) are two essential control parameters. The Weber number, ${{\We}_{g}}={{\rho }_{g}}V_{g}^{2}{{d}_{0}}/\sigma $, represents the ratio of the disrupting aerodynamic force to the restorative surface tension force, where $\rho_g$ and $V_g$ are the density and the relative velocity of the ambient gas, $d_0$ and $\sigma$ are the diameter and the surface tension of the droplet, respectively. The Ohnesorge number, $\Oh={{\mu }_{l}}/\sqrt{{{\rho }_{l}}\sigma {{d}_{0}}}$, represents the relative importance of the viscous force to the inertial and the surface tension forces, where $\mu_l$ and $\rho_l$ are the dynamic viscosity and the density of the droplet, respectively. When the Ohnesorge number is low ($\Oh < 0.1$), with increasing the Weber number, the droplet experiences different breakup modes, such as deformation/vibrational breakup, bag \citep{Kulkarni2014BagBreakup, Opfer2014FilmThickness}, multimode \citep{Cao2007DualbagBreakup, Zhao2013BagStamen}, sheet-thinning \citep{Jain2015SecondaryBreakup, Theofanous2011DropBreakup}, and catastrophic \citep{Huang2018ShockTube, Biasiori2019Highspeed} breakup modes. Since the viscosity of the droplet hinders droplet deformation and induces energy dissipation, the transition Weber number between different modes increases with increasing $\Oh$ \citep{Theofanous2012ViscousLiquids}.

The physical mechanism of droplet breakup is different for different breakup modes. For low $\We_g$ numbers, the Rayleigh-Taylor (RT) instability plays the primary role in the breakup process. The RT instability is an interfacial instability that occurs when a heavy fluid is accelerated to a light fluid. Joseph \emph{et al.}\ \citep{Joseph1999DropBreakup} attributed bag breakup to the relationship between the droplet diameter and the critical wavelength of the RT instability. When the droplet diameter is smaller than the critical wavelength of the RT instability, the fragmentation by the RT instability could not occur. Guildenbecher \emph{et al.}\ \citep{Guildenbecher2009SecondaryAtomization} summarised that the growth of the RT instability resulted in initial surface disturbance, which was strengthened by the aerodynamic effect. Zhao \emph{et al.}\ \citep{Zhao2010BagBreakup} defined an RT wave number ($N_{RT}$) which is the ratio of the maximum cross-stream dimension to the RT instability wavelength (${{N}_{{RT}}}={{d}_{\max }}/{{\lambda }_{\max }}$). Through morphological analysis, they found that the bag breakup occurred when $N_{RT}$ was from $1/\sqrt{3}$ to 1, bag-stamen breakup when $N_{RT}$ was from 1 to 2, and dual-bag breakup when $N_{RT}$ was from 2 to 3. In contrast, for high $\We_g$ numbers, the rim of the droplet cannot form stably due to the weakening of the surface tension, but fragments are directly stripped away from the rim of the droplet. In this condition, the physical mechanism has been debated primarily between two theories, i.e., boundary-layer stripping and sheet thinning. The boundary-layer stripping mechanism, as postulated by Ranger \emph{et al.}\ \citep{Ranger1969Aerodynamic}, is that the ambient airflow would result in the formation of a liquid boundary layer inside the droplet/air interface, and the boundary layer separation of the droplet periphery further leads to the stripping of mass. In contrast, the sheet thinning mechanism, as proposed by Engel \citep{Engel1958SheetThinningMechanism} and Hinze \citep{Hinze1955SheetThinningMechanism}, is that a sheet at the periphery of the droplet forms due to the ambient flow inertia. Subsequently, the sheet breaks into ligaments, and then small droplets are produced from the end of the ligaments.

It should be noted that in the above studies of droplet breakup, the droplets are subjected to uniform flows of air. There are also some studies on the droplet breakup in non-uniform flows. In turbulence flow, due to the non-uniform velocity distribution around the droplet caused by the turbulence velocity fluctuation and turbulence vortices, the deformation characteristic of the droplet is off-center or uneven \citep{Jiao2019TurbulentFlows}. The non-uniform velocity distribution would obviously shift the topology of the droplet breakup only at the large turbulence integral scale (the turbulence vortex scale is close to the droplet size) or the high turbulence intensity. Under moderate turbulence intensity or small-scale turbulent vortices, the turbulence only slightly modifies the breakup morphology of droplets \citep{Omidvar2012TurbulenceFlow, Zhao2019Turbulence}. In disturbed flow fields by the presence of obstacles (solid obstacles or other droplets), the droplet tends to be stretched unilaterally or multilaterally, and results in different breakup modes, such as bullet-like mode \citep{Theofanous2007Aerobreakup} and shuttlecock mode \citep{Stefanitsis2019DropletsInTandem, Stefanitsis2019ClusterDroplet}. The effect of obstacles is significantly affected by the relative position of the obstacle and the droplet \citep{Theofanous2007Aerobreakup, Stefanitsis2019ClusterDroplet, Zhao2019TwoDrops}. Moreover, in continuously accelerated flow, the overall morphology of the droplet changes continuously during the breakup process. Garcia \emph{et al.}\ \citep{Garcia2017AcceleratedFlowField} experimentally found that the deformation of the droplet went through the oblate spheroid shape, hat-like shape, mushroom-like shape, parachute-like shape, and fountain-like shape when the droplet was exposed to a continuously increasing flow.

In many applications, the ambient air for droplet breakup has a strong shear flow. For instance, in a double-swirl combustion chamber of a jet engine, fuel droplets break up under the shear of a double swirl \citep{sankaran2002spray, luo2011direct}. There have been many studies on the fragmentation of droplets in liquids (i.e., liquid-liquid system) due to shear \citep{Guido2011ShearFlow, Barai2016Shear, Ioannou2016Shear, Lanauze2018ElectricFields}. However, the density ratio and the viscosity ratio of the two phases in the liquid-liquid system are both low, different from the high viscosity ratio and the high density ratio of droplet breakup in airflow. K\'{e}kesi \emph{et al.}\ \citep{Kekesi2016ShearFlow} studied the deformation and breakup of a single liquid droplet in shear flow superimposed on a uniform flow. In their numerical simulation, the density ratio was up to 80 and the viscosity ratio was up to 50, but they were still much lower than those in the droplet breakup in shear airflow.

In this experimental study of the deformation and breakup of droplets in shear airflow, we find that the shear effect can not only influence the breakup process remarkably but also lead to a new breakup mode, i.e., the butterfly breakup mode. Then the breakup morphology, breakup mechanism, regime map, and size distribution of secondary droplets are analyzed. The rest of the paper is organized as follows. In Section \ref{sec:sec02}, we describe the experimental setup, including synchronized high-speed imaging and particle image velocimetry (PIV). Then, results are presented and discussed in Section \ref{sec:sec03}. Conclusions are finally drawn in Section \ref{sec:sec04}.

\section{Experimental setup}\label{sec:sec02}
\begin{figure*}
  \centerline{\includegraphics[width=1.3\columnwidth]{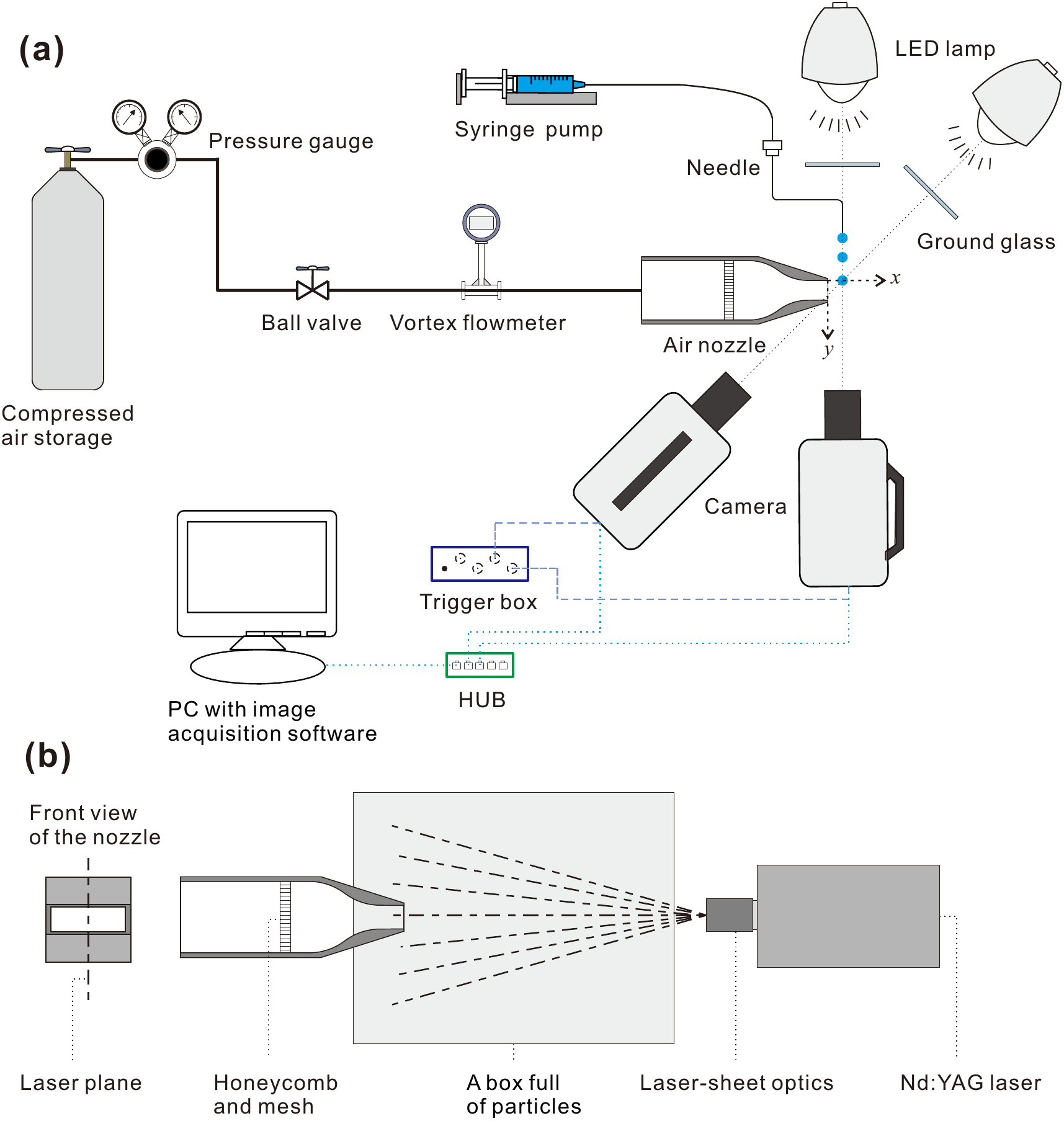}}
  \caption{Schematic diagrams of experimental setup. (a) Experimental setup for high-speed imaging, (b) experimental setup for particle image velocimetry (PIV).}
\label{fig:fig01}
\end{figure*}

In previous experiments of droplet breakup using nozzle jets, droplets often passed through a shear layer of the jet at a fast speed to ignore the influence of the shear layer \citep{Flock2012PIV, Guildenbecher2009SecondaryAtomization, Zhao2010BagBreakup}. In this study, the shear layer of an air jet was used to investigate the deformation and fragmentation of droplets under shear. The experimental setup is shown schematically in FIG.\ \ref{fig:fig01}a. A horizontal rectangular nozzle (the outlet cross-section is 60 mm in width and 20 mm in height) with a compressed-air cylinder was used to generate a continuous air jet. The flow rate was adjusted by a ball valve and measured by a vortex flowmeter (LeiTai LT-LUGB 485, estimated uncertainty $\pm$ 1\%). Experiments were performed in an indoor environment with an air density of $\rho_g$ = 1.185 kg/m$^3$, an air viscosity of $ \mu_g $= 0.0183 mPa$\cdot$s. The Reynolds number of the jet, ${Re}={{\rho }_{g}}V_{g}{{d}_{a}}/\mu_g $, was between 18000 and 23000, where $d_a$ was the hydraulic diameter of the jet, $d_a$ = 30 mm. Therefore, the jet flow was a turbulence flow. According to the studies of Zhao \emph{et al.}\ \citep{Zhao2010BagBreakup, Zhao2019Turbulence}, the turbulence had a small effect on the topology of bag breakup and might only increase the randomness of the droplet breakup process.

Absolute ethanol was used as the droplet fluid, and the density is $\rho_l $= 790 kg/m$^3$, the viscosity is $ \mu_l $= 1.103 mPa$\cdot$s, and the surface tension is $\sigma$ = 21.75 mN/m at room temperature (25 $^\circ$C). Pushed by a syringe pump (Harvard Apparatus, Pump 11 elite Pico plus) at a low speed, droplets formed at the tip of a blunt syringe needle, detached the needle, and entered the air stream under gravity. The size and the speed of the droplet were measured from high-speed images using a customized Matlab program. The diameter of the droplet used in this study is $d_0 = 2.55 \pm 0.05$ mm. The downward speed of the droplet ($V_d$) was in the range of 0.1--1 m/s, which was adjusted by adjusting the falling height of the droplet. Different from the previous studies that the droplets quickly passed through the shear layer, the downward speed of the droplets in this study was low (e.g., $V_d<0.6$ m/s in this study), and the effect of the shear layer was remarkable.

To obtain the details of the droplet breakup process, we used two synchronized high-speed cameras (Photron Fastcam SA1.1) to take images from the side and the bottom views, respectively. The frame rate of the cameras was 5000--8000 fps, and the image resolution was 56--71 $\mu$m per pixel. We used a macro lens with a focal length 60 mm (Nikon AF 60 mm f/2.8D) which has a very low optical distortion. And we set a small aperture (F22 or F16) to obtain a sufficient depth of field which can ensure that most of the fragments can be captured in the images sharply. To ensure sufficient brightness when using the small aperture, we chose high-power (280W) light-emitting diode (LED) lights diffused by ground glass as the background light source.

Particle Image Velocimetry (PIV) experiments were conducted to obtain the velocity field of the shear flow of the air jet, as shown in FIG.\ \ref{fig:fig01}b. Seeding particles were continuously added into the airflow upstream by a Laskin sprayer with dioctyl sebacate as the solution. The particle mist had an average diameter of about 1-3 $\mu$m, which had good fluidity and was stable in the airflow. The particle concentration was adjusted by adjusting the boost pressure of the Laskin sprayer. A large empty box (about $1 \times 1 \times 1.5$ m$^3$) was set at the nozzle exit and was filled with seeding particles before the experiments to obtain the flow field outside the jet. A dual-head Nd:YAG laser was used to illuminate the particles with their second-harmonic output (SOLO120, 532 nm wavelength). The time interval between the two laser pulses was 10 $\mu$s. A CCD camera (Sony ICX085 CCD sensor, Sony Semiconductor Corporation, Fukuoka-shi, Japan) with an image resolution of 1300$\times$1030 pixels and a 42-$\mu$m pixel spacing was used for image acquisition. A PIV system (LaVision) was used to synchronize the camera and the laser, and to post-process the images. In the PIV analysis, the interrogation window started from 64$\times$64 pixels with a 50\% overlap. After adaptive partition, the final vector field corresponded to the interrogation area of 32$\times$32 pixels with a 50\% overlap.

\section{Results and discussion}\label{sec:sec03}
\subsection{Characterisation of shear flow }
\begin{figure*}
  \centerline{\includegraphics[width=1.5\columnwidth]{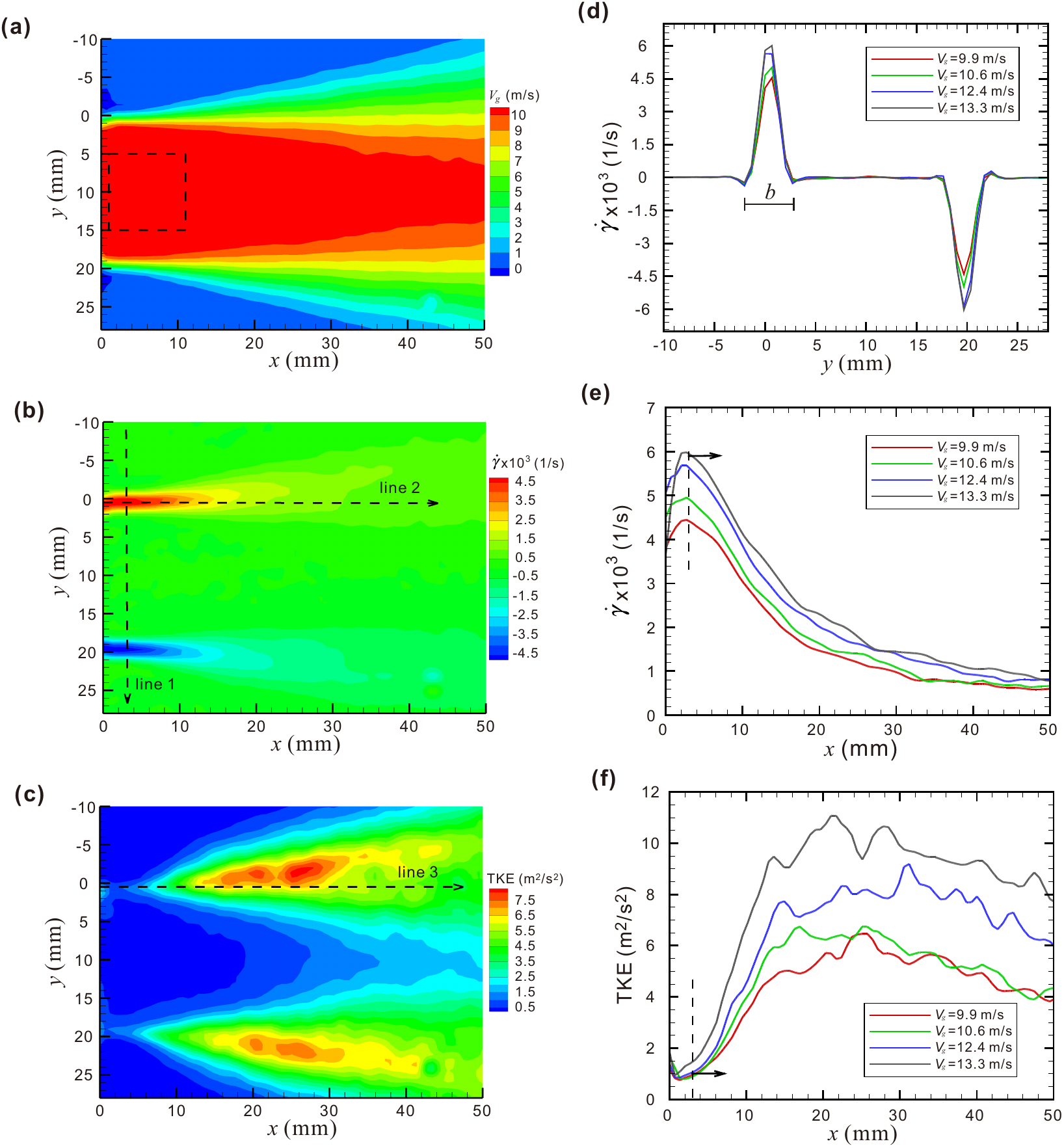}}
  \caption{Velocity field of the shear flow. (a)-(c) Average velocity field, shear strength, and turbulent kinetic energy at $V_g = 10.6$ m/s, (d)-(e) shear strength on line 1 and line 2, (f) turbulent kinetic energy on line 3. The thickness of the shear layer $b$ is labelled in (d), and it is 5 mm in this study even at different air speeds.}
\label{fig:fig02}
\end{figure*}

PIV was used to measure the jet velocity fields at different flow rates. At each flow rate, 250--350 instantaneous velocity fields were obtained, and they were averaged to obtain the average velocity field, as shown in FIG.\ \ref{fig:fig02}a. Then the mean velocity of the jet core (indicated by the rectangle in FIG.\ \ref{fig:fig02}a) was taken as the airflow velocity. From the average velocity field, we can see that the flow can be divided into three regions, i.e., the jet core, the shear layer, and the outer region. There is a large velocity gradient across the shear layer. Therefore, when a droplet passes through the shear layer, the shear layer can induce a strong shear effect on the droplet and affect the breakup process.

To further quantify the velocity fields, we calculated the shear strength ($\dot{\gamma }=\partial {{V}_{x}}/\partial y$, representing the horizontal shear, where $V_x$ is the horizontal component of the velocity vector, as shown in FIG.\ \ref{fig:fig02}b) and the turbulent kinetic energy ($\text{TKE}=\frac{1}{n-1}\sum\nolimits_{i=1}^{n}{{{({{V}_{i}}-{{V}_\text{avg}})}^{2}}}$, representing the variance of the given number ($n$) of vector fields, where $V_i$ and $V_\text{avg}$ are the instantaneous velocity vector and the average velocity vector, as shown in FIG.\ \ref{fig:fig02}c). The number of vector fields ($n$) used to calculate $\text{TKE}$ is 250-350, which is the number of instantaneous velocity fields obtained. When $n>200$, the average velocity field, the shear strength field, and the $\text{TKE}$ field are stable, as shown in FIG.\ S1 in the Supplementary Materials. Representative profiles of the shear strength and the turbulent kinetic energy over a typical droplet trajectory are shown in FIGs.\ \ref{fig:fig02}d-f. As shown in FIGs.\ \ref{fig:fig02}a and \ref{fig:fig02}e, there is a small recirculation zone near the outlet wall ($x<3$ mm). In addition, $\text{TKE}$ begins to decrease after $x = 25$ mm, as shown in FIG.\ \ref{fig:fig02}f, indicating that the entrainment of the jet begins to affect the boundary layer. Therefore, to avoid the influence of the recirculation zone and the entrainment, the entering position of the droplets were all set at 3 mm from the nozzle outlet in the horizontal direction.

\begin{figure}
  \centerline{\includegraphics[width=\columnwidth]{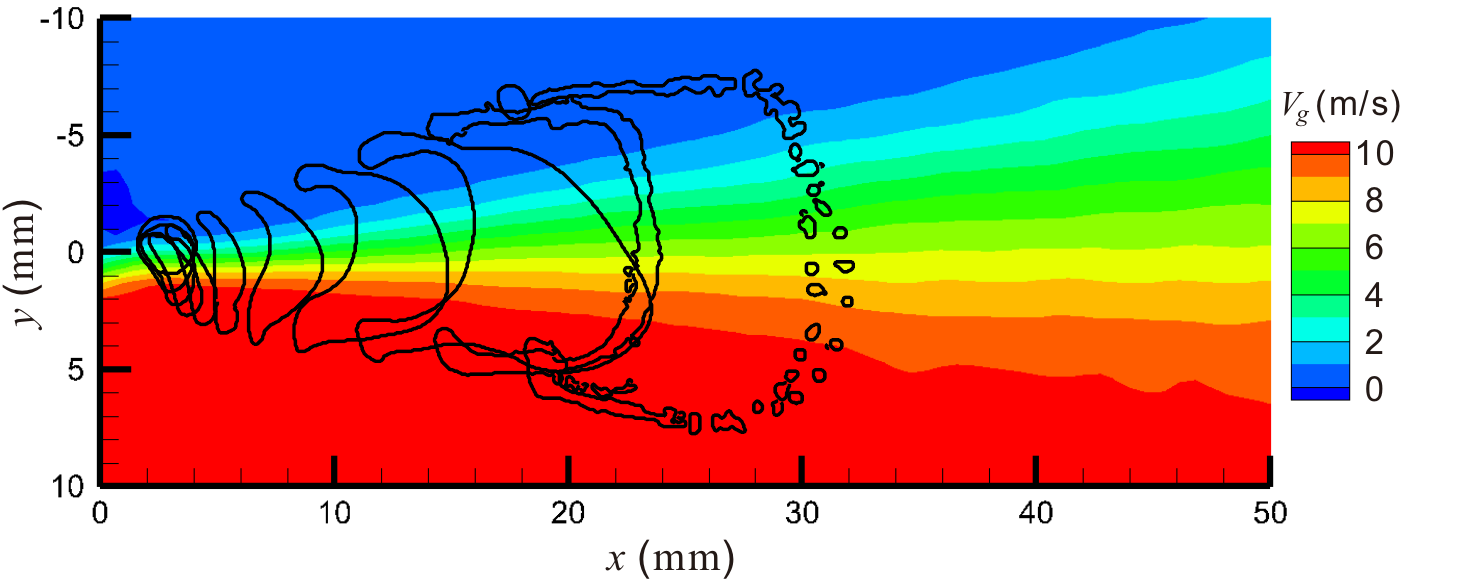}}
  \caption{Contours of the droplet in the shear flow during the breakup process. The time interval between two successive droplet contours is 2 ms. The color contours are the velocity component in the horizontal direction without droplet breakup.}
\label{fig:fig03}
\end{figure}

To understand the flow field around the droplet in the breakup process, we used a customized Matlab program to extract the droplet contours at different time steps, and plotted them in the flow field obtained from PIV, as shown in FIG.\ \ref{fig:fig03}. It can be seen that in the initial deformation stage, the droplet has a low speed in the horizontal direction and stays in the shear layer. After that, with the formation of a bag, the droplet moves downstream, and the shear strength near the droplet decreases. Finally, the droplet breaks up before affected by the jet entrainment. The breakup process of the droplet occurs mainly in the region between 3 mm  $<x<$  25 mm.

\subsection{Breakup morphology in butterfly-breakup mode}
\begin{figure*}
  \centerline{\includegraphics[width=1.4\columnwidth]{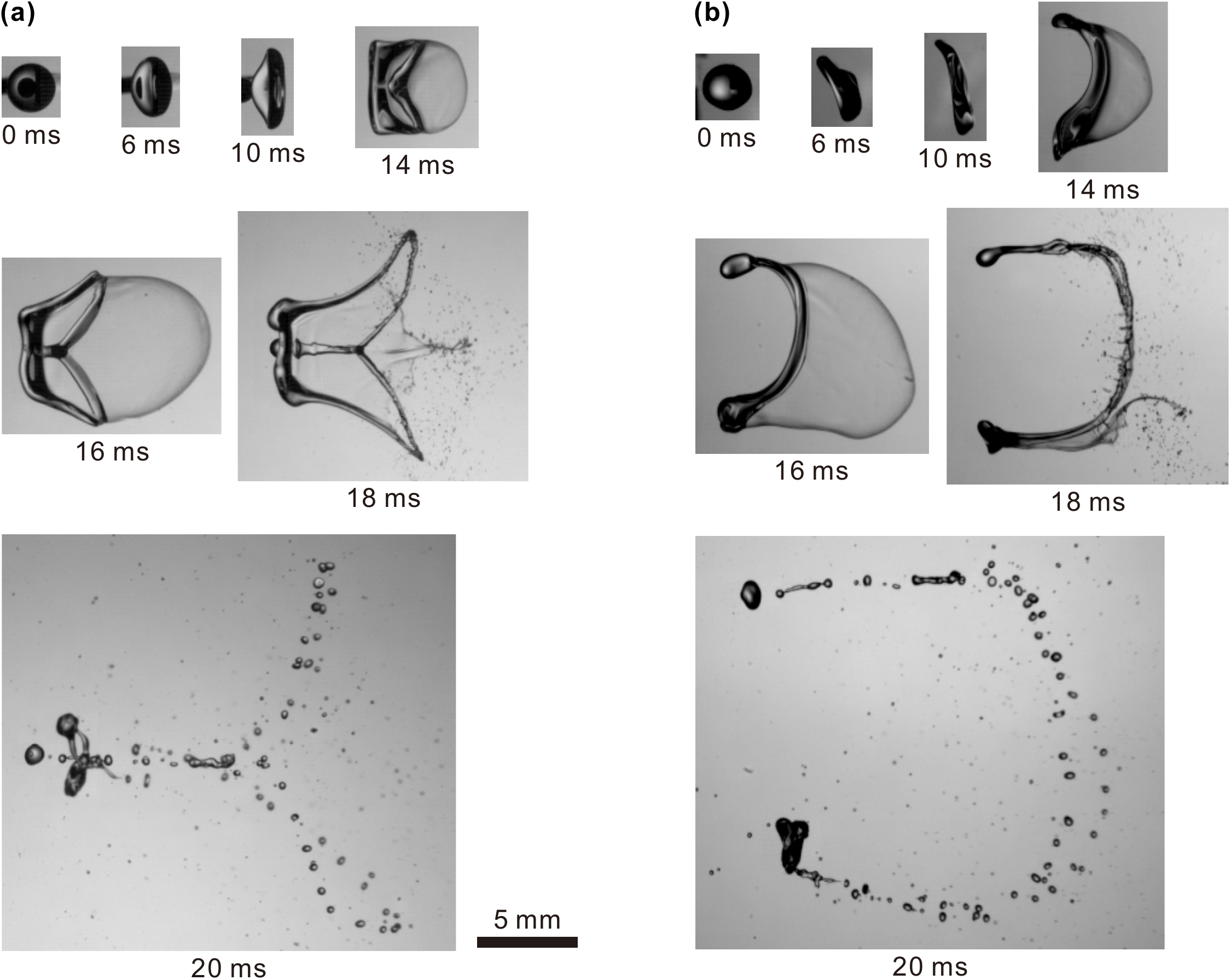}}
  \caption{Image sequences of droplet butterfly breakup in shear flow. The direction of the airflow is from left to right. (a) Bottom view, (b) side view. $\We_d = 3.7$, $\We_g = 14.8$.  The corresponding movie can be found in the Supplementary Materials (movie 1).}
\label{fig:fig04}
\end{figure*}

In shear flows of air, droplets may break up in a butterfly mode, as shown in a typical breakup process in FIG.\ \ref{fig:fig04}. When the droplet enters the shear layer at a low speed, the lower part of the droplet first enters the flow field and is flattened. However, due to the velocity gradient of the airflow, the downward movement of the droplet is decelerated. At the same time, the liquid at the upper part continuously joins the lower part. Therefore, further deformation of the lower part of the droplet is restrained. As indicated by Zhao \emph{et al.}\ \citep{Zhao2013BagStamen} and Opfer \emph{et al.}\ \citep{Opfer2014FilmThickness}, the initial deformation of the droplet by aerodynamic pressure in uniform flow is similar to the collision of a droplet on a wall. Here, the initial deformation in a shear flow is similar to the shape of droplet collision on an inclined wall, i.e., the length of the droplet spreading in the transverse direction is different from that in the longitudinal direction. Viewed from the direction of the airflow, the shape of the droplet is not circular but approximately triangular (10 ms in FIG.\ \ref{fig:fig04}).

In the late stage of the initial deformation, an instability wave begins to develop from the central region of the droplet, and a thin bag attached to the thick ring forms under the action of aerodynamic force (14 ms in FIG.\ \ref{fig:fig04}). In previous studies of droplet breakup in uniform air flows \citep{Zhao2010BagBreakup, Zhao2019Turbulence}, it has been generally accepted that the development of instability waves leads to the formation of nodes on the ring. During the development of the bag in uniform airflow, nodes gradually form and grow due to the growth of the RT instability. However, in this study, due to the shear effect, liquid aggregates at the top and the bottom of the droplets during the initial deformation. The accumulated liquid becomes nodes on the ring at the beginning of the bag development. Then during the development of the bag, due to the difference in liquid mass between the nodes and the inter-node liquid threads, the speed of the nodes moving downstream is slow, while the liquid threads between nodes have a higher speed. Therefore, the ring is continuously stretched by the airflow. Meanwhile, the nodes continue to shrink as the liquid is absorbed by the elongated thread. Finally, the droplet forms a butterfly-shaped structure, as shown at 18 ms in FIG.\ \ref{fig:fig04}. This butterfly-breakup mode can be considered as a variation of the bag breakup mode under the influence of shear.

\subsection{Regime map of droplet breakup}
\begin{figure}
  \centerline{\includegraphics[width=\columnwidth]{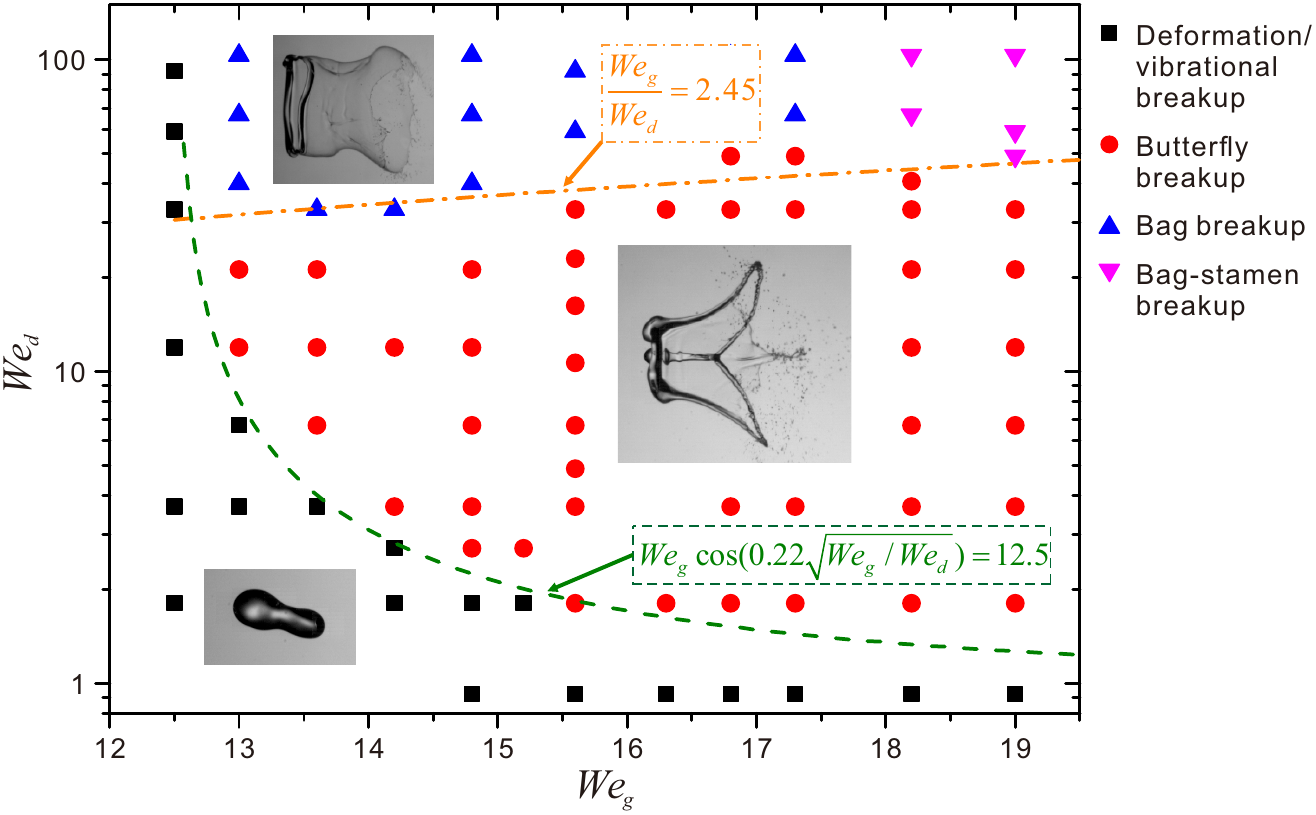}}
  \caption{Regime map of droplet deformation and breakup in shear flows. The orange dash-dotted line indicates the transition from butterfly breakup to bag breakup. The green dashed line indicates the transition from deformation/vibrational breakup to butterfly breakup.}
\label{fig:fig05}
\end{figure}

A regime map of droplet breakup is shown in FIG.\ \ref{fig:fig05}, including the butterfly-breakup mode. We use $\We_d$ to represent the relative magnitude of the downward inertial kinetic energy of the droplet, ${{\We}_{d}}={{\rho }_{l}}{{V}_{d}}^{2}{{d}_{0}}/\sigma $, where $V_d$ is the downward velocity of the droplet when the lower edge of the droplet just reaches the shear layer. When $\We_d$ is very small, due to the lifting effect of the shear layer on the droplet, the droplet does not fully enter the flow field, but only deforms or oscillates. In contrast, when $\We_d$ is very large, the droplet quickly passes through the shear layer, the influence of the shear layer is negligible, and the droplet breakup is the traditional breakup mode in uniform flows. When $\We_d$ is moderate, the process of droplet deformation and breakup is affected by the shear layer, resulting in the butterfly-breakup mode.

To further determine the condition of the butterfly-breakup mode, we introduce a characteristic time for the droplet passing through the shear layer
\begin{equation}\label{eq:eq01}
  {{t}_{y}}=\frac{{{d}_{0}}+b}{{V}_{d}}
\end{equation}
where $b$ is the thickness of the shear layer, $b$ = 5 mm. We also use the timescale of droplet breakup \citep{Jain2019HighDensity}, which is defined as
\begin{equation}\label{eq:eq02}
{{t}^{*}}=\frac{{{d}_{0}}}{{{V}_{g}}}\sqrt{\frac{{{\rho }_{l}}}{{{\rho }_{g}}}}
\end{equation}
Therefore, the ratio of the two timescales ${{t}_{y}}/{{t}^{*}}$ represents the residence timescale of the droplet in the shear layer during the process of droplet deformation. From Eqs.\ (\ref{eq:eq01}) and (\ref{eq:eq02}), we have
\begin{equation}\label{eq:eq03}
{{t}_{y}}/{{t}^{*}}={{C}_{g}}\sqrt{\frac{{\We}_{g}} {{\We}_{d}}}
\end{equation}
where ${{C}_{g}}=({{d}_{0}}+b)/{{d}_{0}}$ is a geometric parameter, and $C_g = 3$ in our experiments. With a decrease in ${{t}_{y}}/{{t}^{*}}$, the shear effect weakens, and the liquid accumulation during the initial deformation reduces. When the liquid accumulated at the top and the bottom is insufficient to form early nodes, i.e., nodes appear randomly instead of fixed at the top and the bottom of the droplet, the butterfly-breakup does not occur. In our experiments, the butterfly-breakup mode disappears when ${{t}_{y}}/{{t}^{*}}<4.7$, corresponding to a transition from butterfly breakup to bag breakup at ${{\We}_{g}}/{{\We}_{d}}=2.45$. This analysis of the boundary agrees well with the experimental data, as shown by the orange dash-dotted line in FIG.\ \ref{fig:fig05}.

\begin{figure}
  \centerline{\includegraphics[width=\columnwidth]{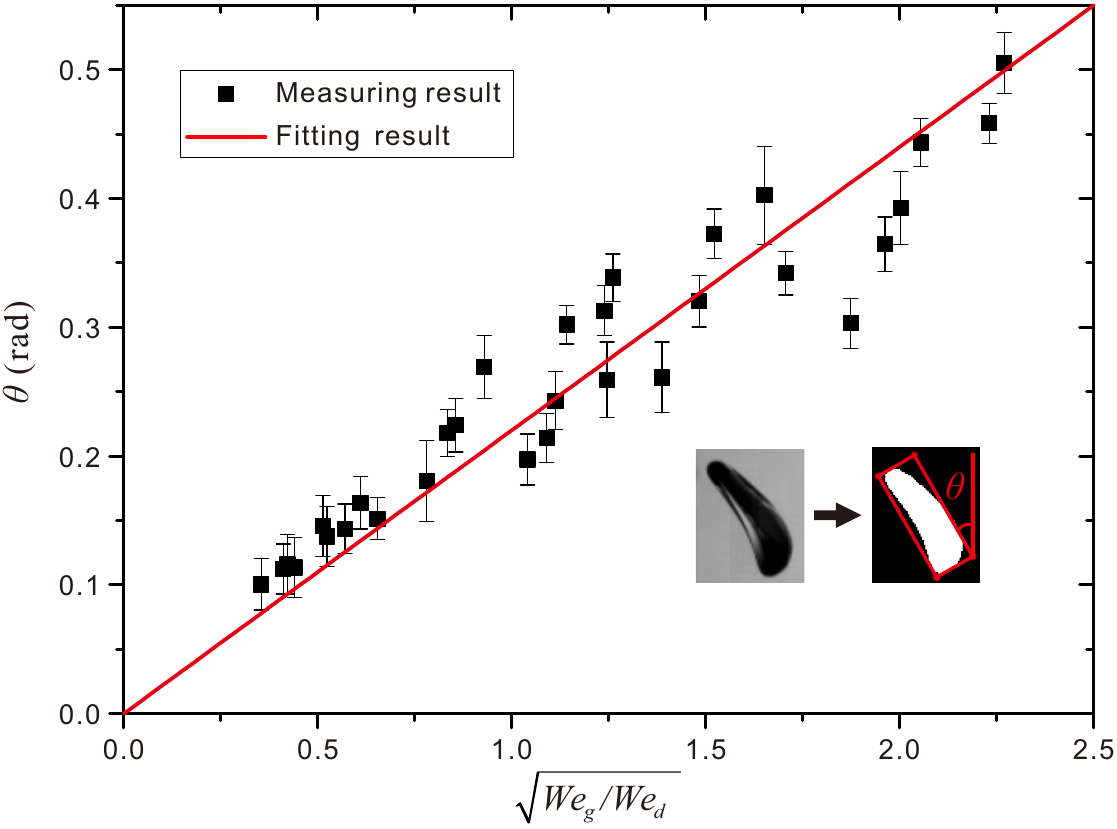}}
  \caption{Deflection angle of the droplet at the instant when the droplet just becomes disc-shaped. The symbols are experimental data, and the error bar of each symbol indicates the standard deviation of about 15 repeated experiments. The line is a least-square fitting, $\theta =0.22\sqrt{{{\We}_{g}}/{{\We}_{d}}}$.}
\label{fig:fig06}
\end{figure}

During the initial deformation of the droplet, the droplet deflects under the shear effect, as shown at 6 ms in FIG.\ \ref{fig:fig04}. The deflection determines the frontal area of the droplet in airflow and remarkably affects the breakup dynamics. The deflection angle when the droplet just becomes disc-shaped was measured from high-speed images via image processing. In the image processing, we determined a smallest bounding box covering the droplet. However, it should be noted that when the tilt angle of the bounding box is different, the aspect ratio of the bounding box is different. Since the droplet was flattened, we chose the tilt angle of the box that has the largest aspect ratio as the deflection angle of the droplet, as shown in FIG.\ \ref{fig:fig06}. By calculating the deflection angle of a large number of droplets, we found the deflection angle of the droplet $\theta$ is approximately proportional to $\sqrt{{{\We}_{g}}/{{\We}_{d}}}$. From a least-square fitting, we have
\begin{equation}\label{eq:eq04}
\theta \approx 0.22\sqrt{\frac{{\We}_{g}}{{\We}_{d}}}.
\end{equation}
The Weber number in airflow can be defined as \citep{Strotos2018CriticalWeberNumber}
\begin{equation}\label{eq:eq05}
{{\We}_{g}} =
\frac{\text{aerodynamic drag force}}{\text{surface tension force}}
=\frac{4{{\rho }_{g}}{{V}_{g}}^{2}{{A}_{f}}}{\pi d\sigma },
\end{equation}
where $A_f$ is the frontal area of the droplet, which is
\begin{equation}\label{eq:eq06}
{{A}_{f}}=\pi {{d}^{^{_{2}}}}/4 .
\end{equation}
Due to the deflection of the droplet, the frontal area of the droplet decreases. The effective frontal area can be approximated as
\begin{equation}\label{eq:eq07}
{{A}_\text{eff}}={{A}_{f}}\cos \theta .
\end{equation}
Therefore, substituting Eqs.\ (\ref{eq:eq04}), (\ref{eq:eq06}), and (\ref{eq:eq07}) into Eq.\ (\ref{eq:eq05}), we can obtain the effective Weber number
\begin{equation}\label{eq:eq08}
{{\We}_\text{eff}}={{\We}_{g}}\cos \left( 0.22\sqrt{\frac{{\We}_{g}}{{\We}_{d}}} \right).
\end{equation}
The deflection of the droplet will lead to a decrease in the effective Weber number. This is the reason why the butterfly-breakup mode still appears even when $\We_g > 18$. In contrast, the breakup in uniform airflows changes from bag mode to bag-stamen mode at about $\We_g = 18$ \citep{Dai2001MultimodeBreakup, Zhao2013BagStamen}. In addition, the transition from deformation/vibrational breakup to bag breakup is the boundary of butterfly-breakup mode. In the uniform flow of air at low $\Oh$ numbers ($\Oh < 0.1$), the transition Weber number between deformation/vibrational breakup and bag breakup is $11 \pm 2$ \citep{Guildenbecher2009SecondaryAtomization, Yang2017TransitionsWeberNumber}. In our experiments, the transition Weber number from deformation/vibrational breakup to butterfly breakup is about 12.5, corresponding to another boundary of the butterfly-breakup mode, ${{\We}_{g}}\cos \left( 0.22\sqrt{{{\We}_{g}}/{{\We}_{d}}} \right)=12.5$. This analysis agrees well with the experimental data, as shown by the green dashed line in FIG.\ \ref{fig:fig05}.

\subsection{Elongation of droplet rim}
\begin{figure}
  \centerline{\includegraphics[width=\columnwidth]{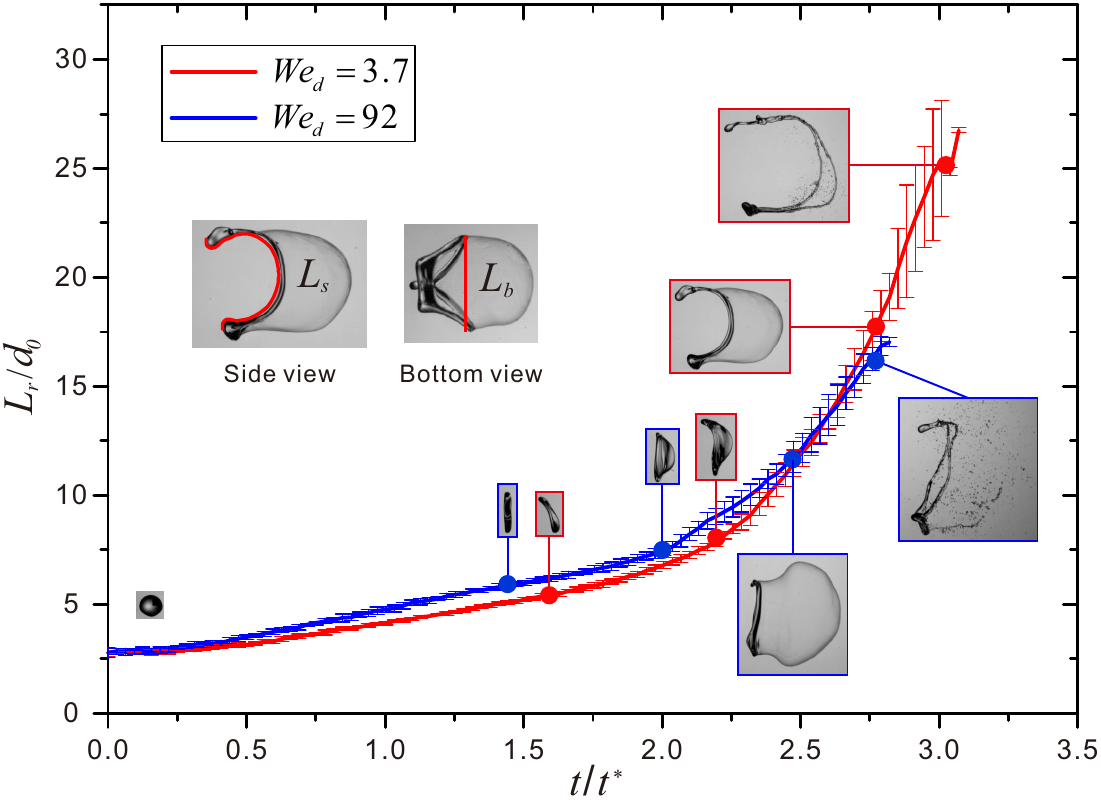}}
  \caption{Length development of the droplet rim from the instant when the droplet edge just enters the shear flow to the end of the bag breakup. $\We_g = 14.8$. The error bars indicate the standard deviations of six repeated processes. The red curve and the images with red frames are for the droplet with intense shear effect ($\We_d = 3.7$), while the blue curve and the images with blue frames are for the droplet with negligible shear effect ($\We_d = 92$). The corresponding movie can be found in the Supplementary Materials (movie 1 for $\We_g = 3.7$ and movie 2 for $\We_d = 92$).}
\label{fig:fig07}
\end{figure}

The length of the droplet rim is an essential parameter of the droplet breakup process, especially when the droplet ring is significantly elongated in the butterfly-breakup mode due to the shear effect. We obtained the rim side length ($L_s$) and the rim bottom width ($L_b$) from the high-speed images via image analysis, and used them to estimate the perimeter of the rim by ${{L}_{r}}=\pi ({{L}_{s}}+{{L}_{b}})/2$ approximately, as shown in FIG.\ \ref{fig:fig07}. To analyze the effect of shear on droplet rim length, we performed different experiments based on dimensionless parameters $\We_d$ and $\We_g$.

Figure \ref{fig:fig07} shows the length development of the droplet rim from the instant that the droplet edge just enters the shear flow to the end of the bag breakup. Since the air flow is turbulence and the breakup process is induced by instability, uncertainties are involved in the experimental results. Therefore, the data presented in this study are from repeated experiments and the standard deviations are included in the plots as error bars. Compared between the cases when the shear effect is negligible ($\We_d = 92$) and intense ($\We_d = 3.7$), the length development of the rim is remarkably different in FIG.\ \ref{fig:fig07}. When the shear effect is intense ($\We_d = 3.7$), during the initial deformation of the droplet, the flattening speed of the droplet is slower and the flattening time is longer. And during the bag development and fragmentation, due to the emergence of the early nodes, the elongation speed of the rim becomes faster and the elongation length becomes longer in the end. Overall, the change of the rim development speed under the effect of strong shear is more dramatic. In contrast, when the shear effect is negligible ($\We_d = 92$), the rim length development is more stable, which means the rim development speed is faster in the stage of the initial deformation and slower in the stage of the bag development and fragmentation than that at strong shear. Finally, due to the different development speeds of the rim in different stages, the two lines representing the intense shear effect and the negligible shear effect will intersect at a point during bag development. At the moment of the intersection, the rims for the two droplets have the same length.

To analyze the effect of shear on droplet deformation in detail, we divide the rim development process into two stages, i.e., the initial deformation stage and the subsequent bag development and fragmentation stage. We set the moment corresponding to the droplet entering the airflow completely as the starting time of the initial deformation stage, i.e., $t_1$. According to the literature \citep{Dai2001MultimodeBreakup, Zhao2013BagStamen}, the initial deformation of the droplet completes when $L_s/d_0$ is about 2, so we choose the moment corresponding to $L_s/d_0  = 2$ as the starting time of bag development, i.e., $t_2$. In addition, we select the instant that the bag is completely fragmented as the end of the second stage.

\begin{figure}
  \centerline{\includegraphics[width=1\columnwidth]{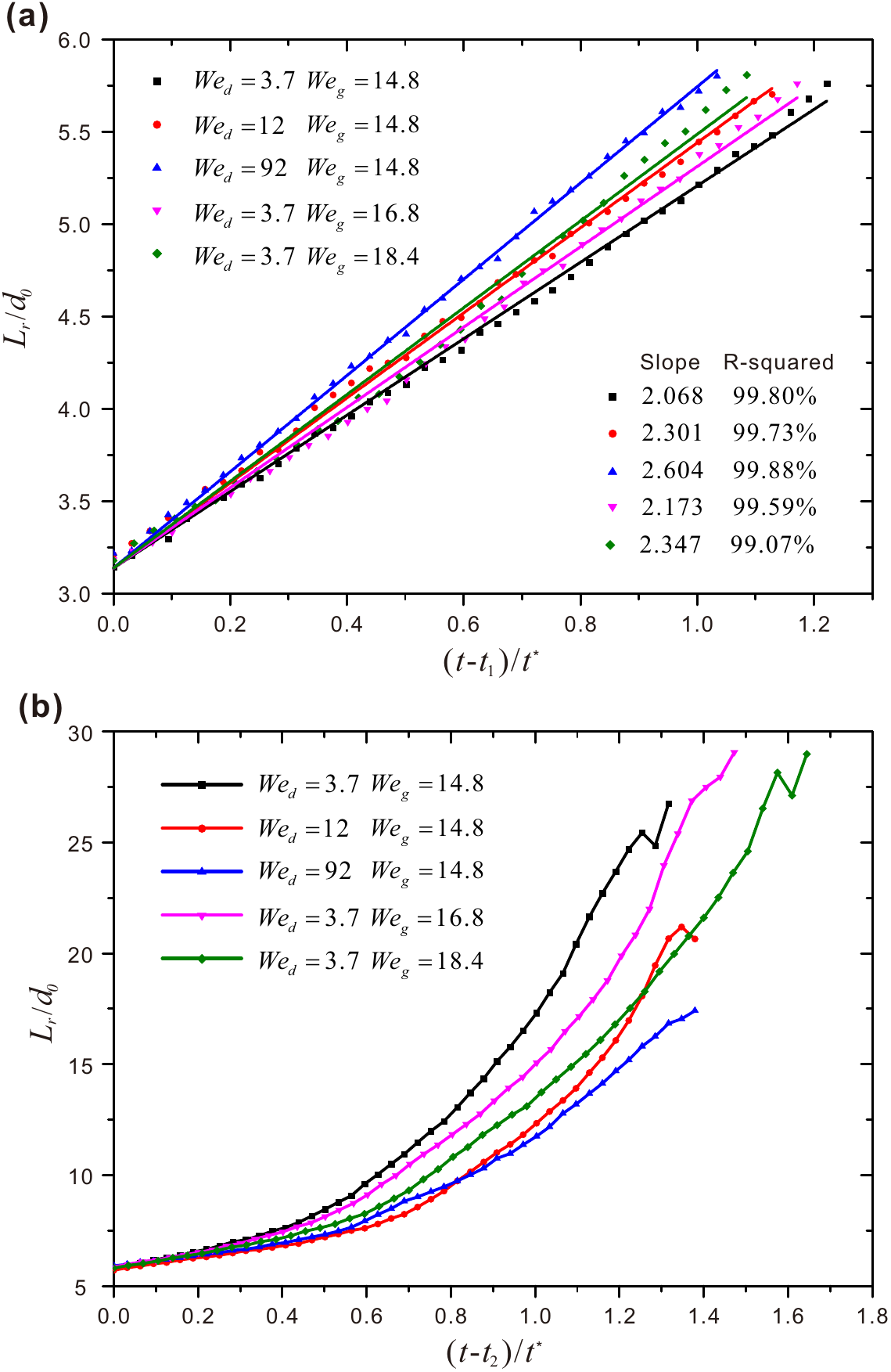}}
  \caption{(a) Rim length in the initial deformation stage. The time is from the droplet entering the airflow completely to the end of the initial deformation. The lines are linear fitting. (b) Rim length in the bag development and fragmentation stage. The time is from the end of initial deformation to the end of bag fragmentation. Each curve shows the average of six repeated experiments.}
\label{fig:fig08}
\end{figure}

The temporal evolution of the rim length during the initial flattening process is shown in FIG.\ \ref{fig:fig08}a. The rim length grows linearly with time, and this is consistent with that in uniform air flows \citep{Cao2007DualbagBreakup, Chou1998SecondaryBreakup, Zhao2013BagStamen}. The flattening is owing to the high-pressure zones formed at the front and the rear stagnation points when air flows around the droplet. In a uniform flow, the stagnation points are at the center of leeward and windward, and the droplet is squashed from the center \citep{Theofanous2011DropBreakup}. However, in a shear flow, the front stagnation point is at the lower part of the droplet. The high-pressure zone locates at the lower part of the droplet as well, which not only causes droplet deflection during the flattening process but also produces a lift force on the droplet. The droplet deflection reduces the effective frontal area, and the lifting effect inhibits the droplet from further penetrating the mainstream, where the airflow has a higher velocity. Both of them can decrease the flattening speed. As shown in FIG.\ \ref{fig:fig08}a, under the same $\We_g$, the flattening speed increases with the increase in $\We_d$ from 3.7 to 92. This is because an increase in $\We_d$ means increasing the downward inertia, which allows the droplet to enter the higher-speed region more quickly under the same lift force. Then the droplet flattens in the region with a higher air velocity, and the flattening speed increases. With the increase in $\We_g$ from 14.8 to 18.4 under the same $\We_d$, the lift force increases, which has a negative effect on the flattening speed. However, the increase in the air velocity due to the increase in $\We_g$ has a greater positive effect on the flattening speed than the suppression effect caused by the increase in the lift force. Therefore, as $\We_g$ increases, the flattening speed of the droplet increases, as shown in FIG.\ \ref{fig:fig08}a.

The temporal evolution of the rim length in the bag development and fragmentation stage is shown in FIG.\ \ref{fig:fig08}b. In the case of $\We_g = 14.8$, as $\We_d$ increases from 3.7 to 92, the development speed of the rim length decreases. This is because for a larger $\We_d$ number, the droplet passes through the shear layer faster, which results in less liquid accumulation during the initial deformation stage, i.e., the rim is more stable. Therefore, the more stable rim has a lower elongation speed and is stretched less. In the case of $\We_d =3.7$, as $\We_g$ increases from 14.8 to 18.4, the development speed of rim length also decreases. This is because when $\We_d$ is small and $\We_g$ is large, due to the lift force of the shear layer, the downward speed of the droplet will reduce to zero and the droplet starts to move upward and departs from the mainstream. Therefore, the development speed of the rim length decreases with the increase in $\We_g$. But meanwhile, the development speed of the rim length is more stable, which will make the rim survives longer. Therefore, as $\We_g$ increases, the rim is longer at the end.

\subsection{Fragmentation of droplet rim}

\begin{figure}
  \centerline{\includegraphics[width=\columnwidth]{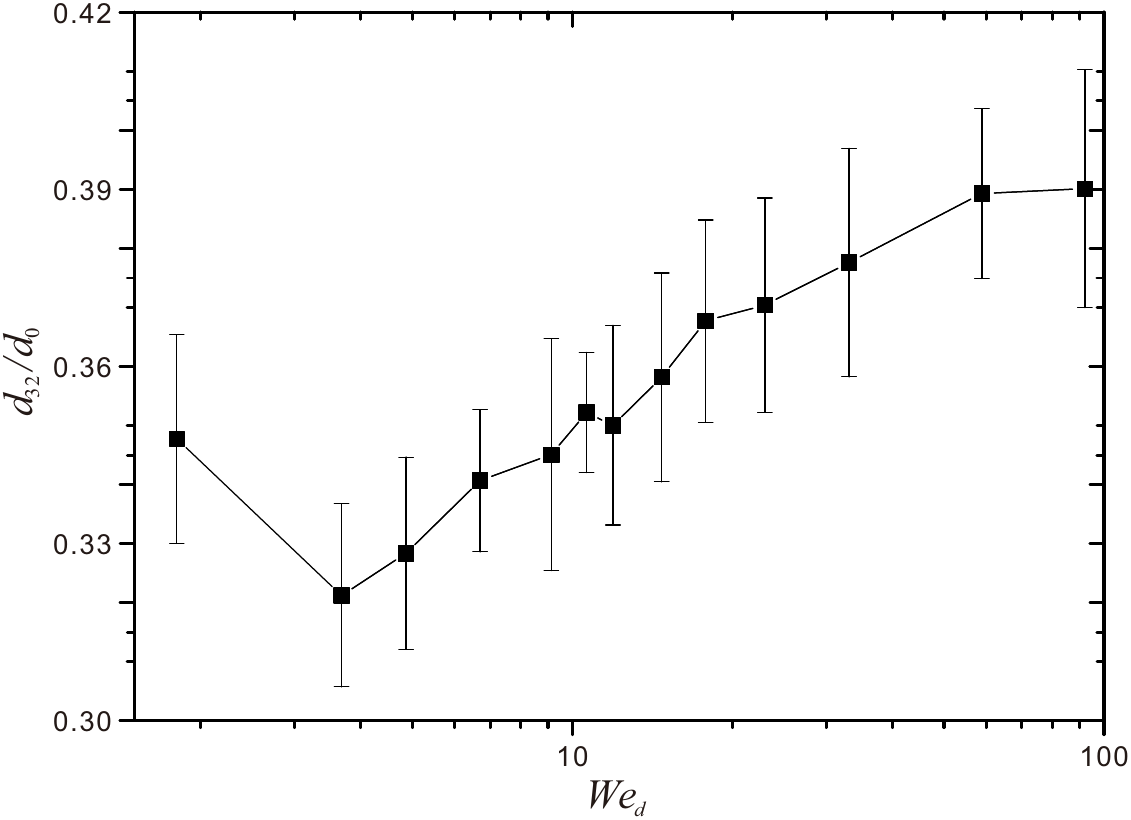}}
  \caption{SMD of rim fragments under different $\We_d$. $\We_g=15.6$. The error bar of each point indicates the standard deviation of about 30 repeated experiments.}
\label{fig:fig09}
\end{figure}

In this section, the shear effect of the airflow on the fragmentation of the droplet rim is analyzed. The Sauter mean diameter (SMD) of the secondary droplets is used in the analysis, which is the most commonly used parameter to evaluate the performance of atomization \citep{Sazhin2014SMD} in many applications of sprays. The SMD is the diameter of droplets in an ideal uniform droplet group which has the same volume/surface area ratio with the actual droplet group, and is defined as ${{d}_{32}}={\sum{{{d}_{s}}^{3}}}/{\sum{{{d}_{s}}^{2}}}$, where $d_s$ is the diameter of secondary droplets. Since the rim contains at least 80\% of the mass of the original droplet \citep{Guildenbecher2017FragmentSize, Zhao2011FragmentSize}, the fragmentation of the rim has a major effect on the atomization performance. Therefore, in this study, we focus on the SMD of rim fragments, which was measured from high-speed images using a customized Matlab program via image processing. The detailed description of the procedure for calculating the droplet diameter has been given in the Supplementary Materials.

\begin{figure*}
  \centerline{\includegraphics[width=1.5\columnwidth]{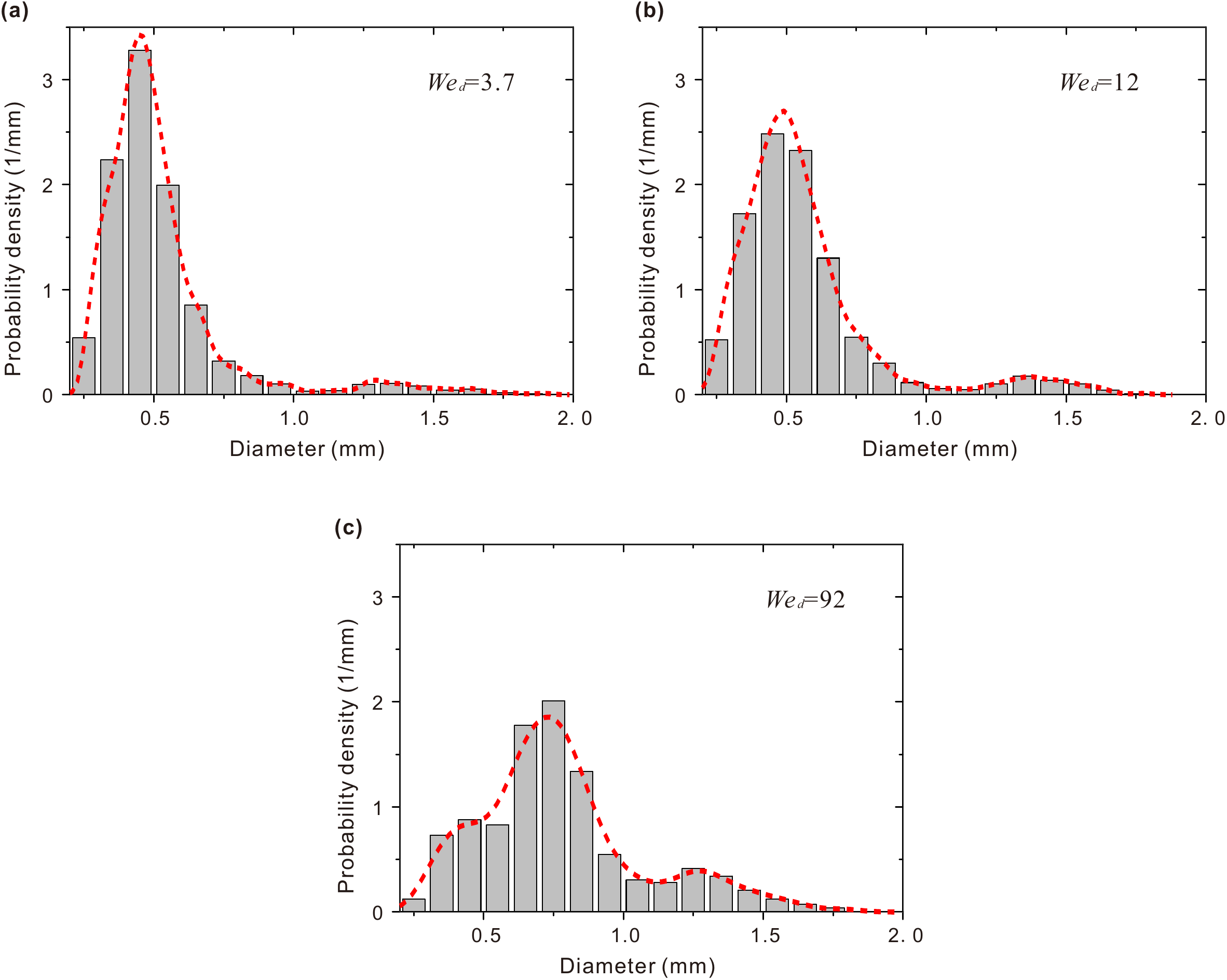}}
  \caption{Size distribution of fragments. (a) $\We_d = 3.7$, (b) $\We_d = 12$, (c) $\We_d = 92$. $\We_g = 15.6$. The dashed curves, obtained using the kernel-based fitting, conform to the bimodal distribution. Each plot shows the statistical results of about 30 repeated experiments.}
\label{fig:fig10}
\end{figure*}

The SMD of rim fragments under different $\We_d$ is shown in FIG.\ \ref{fig:fig09}. As $\We_d$ increases, $d_{32}$ decreases first and then increases. The initial decrease in $d_{32}$  is because when the downward inertia of the droplet is small (i.e., $\We_d$  is small), the droplet is pushed away from the mainstream due to the lift force of the shear layer. Since the droplet does not completely enter the mainstream, the fragments after the breakup are large. When $\We_d > 3.7$, the downward inertia is sufficient to overcome the lift force of the shear layer and shoot the droplet into the mainstream. Therefore, as $\We_d$ increases further, the downward inertia of the droplet increases as well, the droplet enters the mainstream faster and is less affected by the shear layer. This leads to the decrease in liquid accumulation in nodes, i.e., the reduction in the mass distribution inhomogeneity on the droplet rim, which further leads to the decrease in the stretched length of the ring, as shown in FIGs.\ \ref{fig:fig07} and \ref{fig:fig08}b. As a consequence, the diameter of the ring becomes larger, and the size of the fragments generated under capillary instability is larger. Therefore, when $\We_d > 3.7$, the SMD of the rim fragments increases with increasing $\We_d$.

The SMD in uniform flows has been found to be ${{d}_{32}}/{{d}_{0}}=C{\We}_g^{n}$ (for $\Oh < 0.1$), where $C$ is a constant \citep{Hsiang1992SecondaryBreakup, Ng2008BagBreakup, Zhao2011FragmentSize}. In our experiment of shear airflow, we need to use $\We_\text{eff} $ obtained through Eq.\ (\ref{eq:eq08}) instead of using $\We_g$ directly. When $\We_g$ in Eq.\ (\ref{eq:eq08}) is fixed ($\We_g = 15.6$), excluding the cases when the droplets are pushed out of the shear layer ($\We_d < 3.7$), we can obtain a fitting correlation about the relationship between the $d_{32}$ and $\We_d$
\begin{equation}\label{eq:eq09}
  \frac{{{d}_{32}}}{{{d}_{0}}}=C{{\We}_\text{eff}}^{n}=0.383\cos {{(0.87{{\We}_{d}}^{-0.5})}^{2}}.
\end{equation}
When $\We_d$ is infinite, the influence of the shear is negligible and Eq.\ (\ref{eq:eq09}) approaches a constant $d_{32}/d_0 = 0.383$, which is close to the experimental results of Chou and Faeth \citep{Chou1998SecondaryBreakup} ($d_{32}/d_0 = 0.36$ at $\We_g = 15$). In addition, the experimental result of Zhao \emph{et al.}\ \citep{Zhao2013BagStamen} showed that the average size of the ring and stamen was $d_{32}/d_0 = 0.31$ in the range of $16 < \We_g < 26$.

The size distribution of the fragments under different $\We_d$ is compared in FIG.\ \ref{fig:fig10} and exhibits two distinct peaks, i.e., a bimodal distribution. The first peak corresponds to the ring breakup, while the second peak corresponds to the node breakup. As the ring is stretched longer (i.e., $\We_d$ decreases), the first peak is higher, and the corresponding size of the fragments is smaller. In addition, because the liquid accumulated at the nodes is continuously absorbed during the elongation of the ring, the size of the fragments corresponding to the second peak does not increase significantly.

\begin{figure}
  \centerline{\includegraphics[width=0.9\columnwidth]{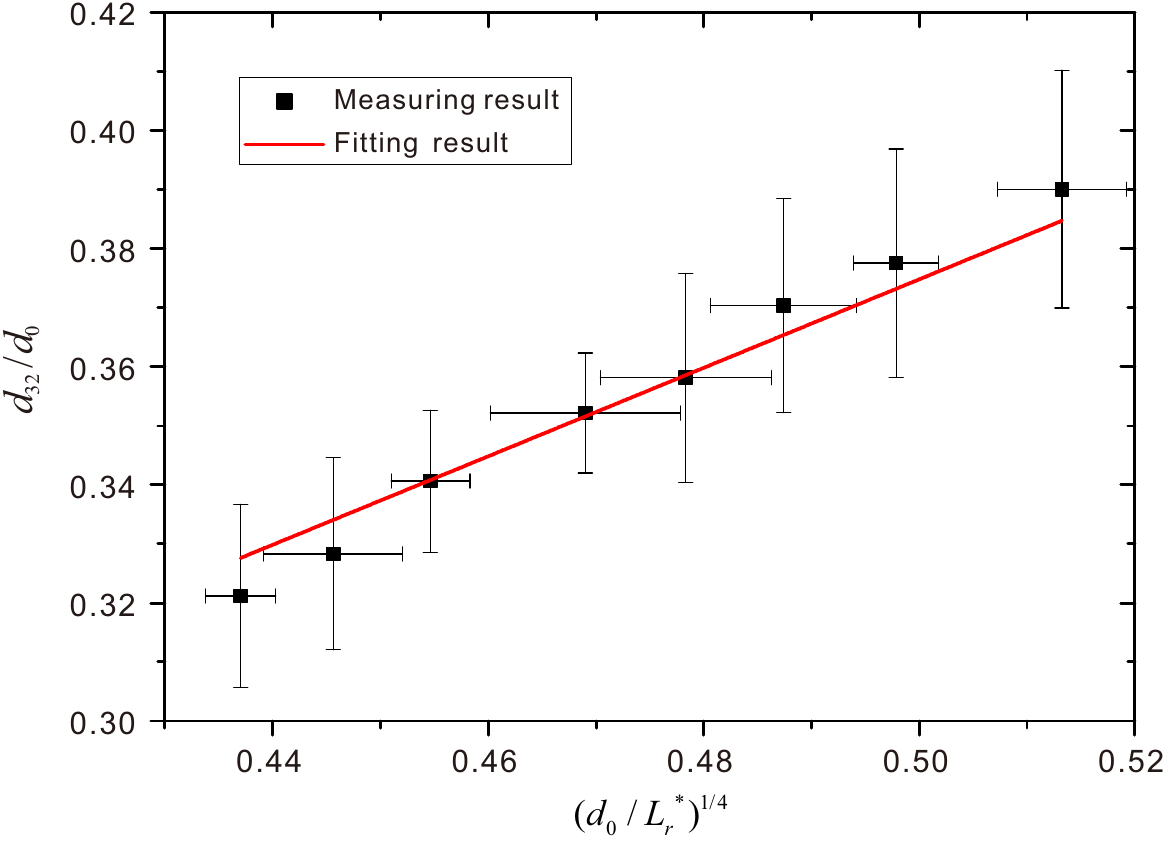}}
  \caption{Relationship between the Sauter mean diameter ($d_{32}$) and the maximum rim length, $L_r^*$. $\We_g = 15.6$. The red line is a linear fitting using Eq.\ (\ref{eq:eq13}) with $C_r = 0.2$. The error bars indicate the standard deviations. For each point, the error bar of  $({d_0}/{L_r^*})^{1/4}$ indicates the standard deviation of about 15 repeated experiments, and the error bar of $d_{32}/d_0$ indicates the standard deviation of about 30 repeated experiments.}
\label{fig:fig11}
\end{figure}

To further explore the relationship between the length of the rim and the size of the rim fragments, we need to understand the mechanism of rim breakup. The rim breakup is the capillary breakup of a liquid column in a crossflow, which is resulted from the column wave growth on the surface \citep{Madabhushi2006LiquidJet}. At a low viscosity, the surface tension causes the rim to form droplets, which will detach when the aerodynamic drag force exerted by the airflow exceeds the surface tension force. Thus, the size of the fragments can be estimated from this force balance
\begin{equation}\label{eq:eq10}
{{\rho }_{g}}{{V}_{g}}^{2}\pi {{d}_{32}}^{2}\sim \pi \sigma {{d}_{r}},
\end{equation}
where $d_r$ is the diameter of the rim. For simplicity, we assume that the rim has a uniform diameter, and ignore the influence of the node breakup. Since the rim accounts for most of the mass of the original droplet \citep{Guildenbecher2017FragmentSize, Zhao2011FragmentSize}, we assume that
\begin{equation}\label{eq:eq11}
\frac{\pi {{d}_{r}}^{2}}{4}{L}_{r}^{*}=\frac{\pi {{d}_{0}}^{3}}{6},
\end{equation}
where $L_r^*$ is the length of the rim at the instant of the bag to be completely fragmented. The existence of the bag has a stabilising effect on the rim. The development of the capillary instability on the rim begins after the bag is completely fragmented. Therefore, the rim diameter at the moment of the complete bag fragmentation is the initial rim diameter for the capillary instability. So $L_r^*$ can be regarded as the maximum length before the droplet rim starts to break. Through Eq.\ (\ref{eq:eq11}), we can obtain the rim diameter
\begin{equation}\label{eq:eq12}
{{d}_{r}}=\sqrt{\frac{2{{d}_{0}}^{3}}{3{L}_{r}^{*}}}.
\end{equation}
Finally, substituting Eq.\ (\ref{eq:eq12}) into Eq.\ (\ref{eq:eq10}), and then rearranging Eq.\ (\ref{eq:eq10}), we can get the relationship between the Sauter mean diameter ($d_{32}$) and the maximum rim length ($L_r^*$)
\begin{equation}\label{eq:eq13}
\frac{{{d}_{32}}}{{{d}_{0}}}={{C}_{r}}{{\We}_{g}}^{1/2}{{(\frac{{{d}_{0}}}{{L}_{r}^{*}})}^{1/4}},
\end{equation}
where $C_r$ is a constant. The experimental data of this scaling are shown in FIG.\ \ref{fig:fig11}. With increasing $L_r^*$, the $d_{32}$ decreases. When $\We_g = 15.6$, the constant in Eq.\ (\ref{eq:eq13}) is $C_r = 0.2$ according to the linear fitting of the results.

\section{Conclusions}\label{sec:sec04}
In this study, we experimentally investigate the shear effect on the breakup of droplets in air flows, and identify a new mode of droplet breakup, i.e., the butterfly-breakup mode. The shear effect is generated by the shear layer of an air jet, and is characterized by the velocity field, the shear strength, and the turbulent kinetic energy. During the butterfly breakup, the lower part of the droplet first enters the shear layer and experiences a strong aerodynamic pressure, which deflects the droplet and results in a butterfly-shaped bag. A regime map of droplet breakup is produced in $\We_d$--$\We_g$ space, and the transitions between different modes are obtained based on scaling analysis, which are ${{\We}_{g}}\cos \left( 0.22\sqrt{{{\We}_{g}}/{{\We}_{d}}} \right)=12.5$ for the transition from the deformation/vibrational-breakup mode to the butterfly-breakup mode and ${{\We}_{g}}/{{\We}_{d}}=2.45$ for the transition from the butterfly-breakup mode to the bag-breakup mode, respectively. The elongation of the droplet rim is affected by the shear effect. During the initial deformation of the droplet, the flattening of the droplet in the shear flow is slower than that in uniform flow, while during the bag development and fragmentation, the rim is significantly elongated by shear. The fragmentation of the droplet rim is also affected by the shear effect. As $\We_d$ increases, the Sauter mean diameter of the secondary droplets decreases first because of the lift effect of the shear layer, and then increases with $\We_d$ because the droplet can rapidly enter the mainstream and the inhomogeneity on the droplet rim is reduced.

The deformation and aero-breakup of droplets in shear flows of air is a complex process. This study mainly focuses on the butterfly-breakup mode due to the shear effect. There are many open questions in this area yet to be answered, such as the effect of fluid properties, the effect of environmental pressure and temperature, and the detailed flow fields during the breakup process. Further studies in this area will not only provide physical insight into this fundamental process, but also be useful for the design and optimization of relevant applications.

\section*{Supplementary Materials}
See supplementary materials for movies of droplet deformation and breakup in airflow with intense (Movie 1, $We_d=3.7, We_g=14.8$) and weak (Movie 2, $We_d=92, We_g=14.8$)  shear effects, the effect of the number of instantaneous flow fields on the measurement (Section S1 and FIG.\ S1), and the procedure for calculating the fragment diameter (Section S2 and FIG.\ S2).

\section*{Acknowledgements}
This work is supported by the National Natural Science Foundation of China (Grant no.\ 51676137) and the National Science Fund for Distinguished Young Scholars (Grant no.\ 51525603).

\section*{Data Availability Statement}
The data that support the findings of this study are available from the corresponding author upon reasonable request.

\bibliography{dropletBreakup}
\end{document}